\documentclass[useAMS,usenatbib]{mn2e}

\usepackage{amssymb}
\usepackage{amsfonts}
\usepackage{mncite}
\usepackage{epsfig}
\usepackage{psfig}

\defcitealias{scheuer96}{SF}

\begin{document}

\title[Formation of 2M1207B]{Constraints on the
  formation mechanism of the planetary mass companion of 2MASS
  1207334-393254}

\author[G. Lodato, E. Delgado-Donate and C. J. Clarke]
{G. Lodato$^1$, E. Delgado-Donate$^2$ and C. J. Clarke$^1$\\
 $^1$ Institute of Astronomy, Madingley Road, Cambridge, CB3 0HA\\
 $^2$ Stockholm Observatory, SCFAB, 106 91 Stockholm, Sweden}

\maketitle

\begin{abstract}
  
  In this paper we discuss the nature and the possible formation
  scenarios of the companion of the brown dwarf 2MASS 1207334-393254.
  We initially discuss the basic physical properties of this object and
  conclude that, although from its absolute mass ($5M_{\rm Jup}$), it
  is a planetary object, in terms of its mass ratio $q$ and of its
  separation $a$ with respect to the primary brown dwarf, it is
  consistent with the statistical properties of binaries with higher
  primary mass. We then explore the possible formation mechanism for
  this object. We show that the standard planet formation mechanism of
  core accretion is far too slow to form this object within 10 Myr, the
  observed age of the system. On the other hand, the alternative
  mechanism of gravitational instability (proposed both in the context
  of planet and of binary formation) may, in principle, work and form
  a system with the observed properties.

\end{abstract}


\section{Introduction}

The system 2MASS 1207334-393254 (2M1207) is remarkable.  Found in the
TW Hydra association (distance $\sim 70$ pc, age $\sim 10$ Myr), the
most massive object, 2M1207A, is a brown dwarf with mass
$M_{\star}=25M_{\rm Jup}$ \citep{gizis02}, known to be surrounded by a
circumstellar disc \citep{sterzik04}. The other body, 2M1207B, has a
mass in the planetary range ($M_{\rm s}\sim 5M_{\rm Jup}$) and lies at
a projected distance $R_{\rm s}\approx 55$ AU from 2M1207A
\citep{chauvin04}. This planetary-mass object has been shown to be
comoving with 2M1207A \citep{chauvin05}. We will refer to its
components, 2M1207A and 2M1207B, as primary and secondary
respectively. The intriguing feature of the system is the mass of the
secondary, although we note that mass estimates for these very low mass
objects are still subject to considerable uncertainties. The very
existence of this system raises the question of the possibility of the
formation of planets around brown dwarfs.

However, it is worth noting that this system is quite peculiar, if
considered as a star+planet system. First of all the mass ratio between
the ``star'' and the ``planet'' ($q=M_{\star}/M_{\rm s}=0.2$) is very
high for a planet companion. Secondly, its semi-major axis is quite
large (larger than the orbit of Neptune, which is 30 AU). On the other
hand, this system could be easily considered as a very low mass
binary. Mass ratios of $q\approx 0.2$ (even if they are low in the
context of binaries) are not uncommon in binaries with a solar-like
primary. 

In this paper we discuss whether this system should be considered as a
planetary system or rather as a very low mass binary. In Section 2 we
start by discussing the properties of the system in the context of the
statistical properties of binaries with higher mass primary. Then, in
Section 3, we consider the constraints that the observed properties of
the system place on its formation mechanism. We consider a number of
possible scenarios: the core accretion model (which is generally
consider for planet formation), the gravitational instability model
(considered both for planets and for binary systems) and other
possible binary formation processes. We conclude that the core
accretion model is not able to account for the formation of the
companion of 2M1207, its formation time-scale largely exceeding the
age of the system. On the other hand, the gravitational instability
model and, possibly, other binary formation mechanisms are in
principle able to form this system. 

\section{Is 2M1207 a ``binary-like'' or a ``planet-like'' system?}

Evidently, the absolute mass of the companion of 2M1207 places it
firmly in the planetary regime. Indeed, $5M_{\rm J}$ planets are
commonly detected around solar mass stars; since radial velocity
surveys only probe systems with separation $a\lesssim 5$ AU (i.e.,
$\approx 0.1$ times the separation of 2M1207), we however do not know
whether such massive planets at large separations are common around
solar mass stars.

On the other hand, the {\it mass ratio} is binary-like. In order to
assess this, we compare the properties of 2M1207 with those of
binaries with higher mass primary. In particular, we consider the two
samples of \citet{fischer92}, who analyzed the binary properties of M
dwarfs (with primary mass in the range $0.1-0.57M_{\odot}$), and the
sample of \citet{close03}, who have collected a number of binaries
with primary masses in the range $0.05-0.095M_{\odot}$.

\begin{figure}
\centerline{\epsfig{figure=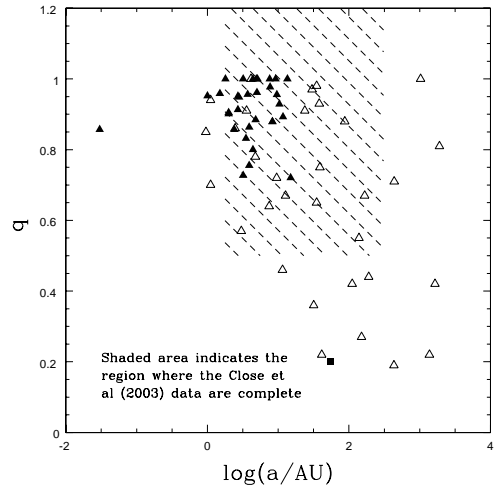,width=0.38\textwidth}}
\caption{Correlation between the logarithm of the separation $a$ (in
AU) and the mass ratio for the binaries observed by Fischer and Marcy
(1992, open triangles) and Close et al. (2003, filled triangles). The
filled square illustrate the properties of the companion in 2M1207.}
\label{fig:samples}
\end{figure}

In Fig. \ref{fig:samples} we show the correlation between separation
and mass ratio for the two samples of \citet{fischer92} (open
triangles) and of \citet{close03} (closed triangles), including only
resolved binaries, for which the mass ratio can be reliably determined
(see discussion in \citealt{mazeh92}). In the \citet{fischer92}
sample, a slight tendency of having lower $q$ for wider systems can be
seen, in agreement with the conclusion of \citet{mazeh92} for higher
mass primaries. On the other hand, the \citet{close03} sample shows a
shortage of systems with large separation, even if very few systems
with separations larger than 10 AU have also been found
\citep{luhman04,phanbao05}. 

Where does 2M1207 stand in this context? The mass ratio and separation
of 2M1207 is plotted in Fig. \ref{fig:samples} with a filled square.
With respect to the \citet{close03} sample, it appears to be unusual in
that it has a larger separation (roughly a factor 3 larger than the
widest system in \citealt{close03}). However, it can be seen that its
properties are perfectly consistent with the distribution seen in
\citet{fischer92}. 2M1207 could therefore be just a rare wide system
for brown dwarf binaries, but with a mass ratio perfectly consistent
with the distribution observed for primaries of higher masses. Note
that the mass of the companion of 2M1207 is below the completeness
level of \citet{close03} and could not have been detected by them.

We therefore conclude that, if the companion orbited a star, it would
have been classified as a planet according to its absolute mass and as
a binary according to its mass ratio $q$ and separation $a$. 

\section{Possible formation mechanisms}

We now proceed to the discussion of the formation mechanisms for this
systems, considering: ({\it i}) the core accretion mechanism
\citep{pollack96}, which is usually considered for planet formation,
({\it ii}) a gravitational instability in a massive disc
\citep{boss00}, which is sometimes invoked both for planets and for
binaries and ({\it iii}) other binary formation mechanisms. 

\subsection{Formation via core accretion}

In this Section we will show that the standard core accretion model,
generally assumed to be the most likely formation mechanism for
planets around solar-like stars, is not able to account for the
formation of the companion of 2M1207 within 10 Myrs, the observed age
of the system.

We will not provide a detailed core accretion model for the formation
of this system. Rather, we will give simple estimates of the relevant
time-scales involved and will try to put constraints on the physical
properties of the protostellar disc where the planet was born, based on
these time-scales. 

We start by noting that if the secondary formed out of a protostellar
disc, this disc must have been quite massive, with $M_{\rm disc}\gtrsim
qM_{\star}$. Such massive discs are uncommon around T Tauri stars, but
might have been more common in earlier phases of star formation. In any
event, such a massive disc must have been self-gravitating. In the
context of gravo-turbulent models of cluster formation, it is difficult
to explain the presence of extended discs surrounding such low mass
objects. In fact, in most numerical simulations of star formation
\citep{bate03}, brown dwarfs are ejected from their parental cloud, and
only in a few cases they are able to retain a large disc after
ejection. On the other hand, there is now evidence for discs on small
scales surrounding rather evolved (T Tauri-like) young brown dwarfs
\citep{klein03,sterzik04}.

The core accretion model assumes that the formation of a giant planet
proceed in two steps. First, planetesimal of mass $m\approx 10^{18}$ g
accrete onto proto-planetary cores. After an initial rapid increase in
the core mass due to accretion, the protostellar disc becomes depleted
of solid material and the planetesimal accretion rate onto the core
decreases substantially. In this phase, the core mass increases much
slowly with time and as the energy input from planetesimal accretion is
turned off, the planetary atmosphere becomes unstable, starting a rapid
phase of gas accretion, which occurs on the Kelvin-Helmholtz
time-scale, $\tau_{\rm KH}$, given by \citep{ida04}:
\begin{equation}
\tau_{\rm KH}=10^9\left(\frac{M_{\rm p}}{M_{\oplus}}\right)^{-3}
\left(\frac{\kappa_{\rm d}}{\mbox{1~g}^{-1}\mbox{cm}^{2}}\right)\mbox{yr},
\label{eq:kh}
\end{equation} 
where $M_{\rm p}$ is the mass of the planetary core and $\kappa_{\rm
d}$ is the opacity of the gas disc. If we assume that the value of the
opacity is the typical interstellar value of 1 cm$^{2}$/g, in order to
be able to accrete the gaseous envelope within 10 Myrs (the age of the
system), the planetary core in our case must therefore reach a mass of
at least $4.6M_{\oplus}$ (see also \citealt{ida04}). Recently,
\citet{hubickyj05} have shown how reducing the opacity can speed-up the
formation process. This is also clear from Eq. (\ref{eq:kh}). However,
since the dependence of the Kelvin-Helmholtz time-scale on mass is
rather steep, even relatively large changes in the opacity only result
in minor modifications of the required core mass. For example, if we
assume $\kappa_{\rm d}=0.02$ cm$^{2}$/g (the value used by
\citealt{hubickyj05}), then the required core mass is only reduced to
$\approx 1.26 M_{\oplus}$.

In order to proceed further, we have to make some simplifying
assumptions. In particular, we will initially assume that the planet
formed at $R_{\rm s}=55$ AU (the effect of planetary migration is
discussed below). We will also assume that the protostellar disc has a
gas density profile $\Sigma_{\rm g}\propto R^{-1}$ and that the
solid-to-gas ratio is 0.01, so that the surface density of solids is
$\Sigma_{\rm d}=0.01\Sigma_{\rm g}$. The surface density of the disc at
$R_{\rm s}$ is:
\begin{equation}
\Sigma_{\rm g}(R_{\rm s})=\frac{M_{\rm disc}}{2\pi R_{\rm s}^2} 
\frac{R_{\rm s}}{R_{\rm out}}\lesssim \frac{M_{\rm disc}}{2\pi R_{\rm
s}^2},
\end{equation} 
where $R_{\rm out}\gtrsim R_{\rm s}$ is the disc outer radius. If we
take $M_{\rm disc}=5M_{\rm Jup}$ (the minimum mass it should have in
order to form the planet) and $R_{\rm s}=55$ AU, we get $\Sigma_{\rm
g}(R_{\rm s})\lesssim 2.33~\mbox{g/cm}^2$. Consequently, the surface
density of solids at $R_{\rm s}$ is $\Sigma_{\rm d}\lesssim
0.022~\mbox{g/cm}^2$.

Note that, for such a density distribution of solids, there is in
principle more than enough solid material in the disc for the core to
grow up to $M_{\rm p}\approx 4M_{\oplus}$. However, the question is
what is the time-scale needed for the core to reach this mass? Here, we
again follow \citet{ida04}, and assume that the core mass as a function
of time is given by:
\begin{eqnarray}
\nonumber M_{\rm p}(t) & \approx & 8\left(\frac{t}{10^6\mbox{yr}}\right)^3
\left(\frac{\Sigma_{\rm d}}{10~ \mbox{g cm}^{-2}}\right)^{21/5}\\
 && \left(\frac{R_{\rm s}}{1\mbox{AU}}\right)^{-9/5}
\left(\frac{M_{\star}}{M_{\odot}}\right)^{1/2} 
\left(\frac{m}{10^{18}\mbox{g}}\right)^{-2/5}
M_{\oplus},
\label{eq:coreacc}
\end{eqnarray}
where $m$ is the mass of the planetesimals and where we have also
assumed that $\Sigma_{\rm d}=0.01\Sigma_{\rm g}$.

Within 10 Myrs (and assuming $M_{\star}=25 M_{\rm Jup}$, $m=10^{18}$ g
and $R_{\rm s}=55$ AU), the mass of the core will therefore be:

\begin{equation}
M_{\rm p}\approx 0.9 M_{\oplus}
\left(\frac{\Sigma_{\rm d}}{10~ \mbox{g cm}^{-2}}\right)^{21/5}.
\end{equation}
In order for $M_{\rm p}$ to be at least $4M_{\oplus}$, the surface
density of planetesimals must be $\Sigma_{\rm d}\gtrsim 14~ \mbox{g
cm}^{-2}$, much larger than the estimate given above ($\sim
0.02~\mbox{g cm}^{-2}$), based on the estimated mass of the disc. If
the density of solids were as high as $14~\mbox{g cm}^{-2}$, then the
total disc mass must have been at least $M_{\rm disc}\approx 3500M_{\rm
Jup}$, which is far too large. Increasing the solid-to-gas ratio to 0.1
would not ease the situation, since it would require a disc mass of
$350M_{\rm Jup}$, which is still unreasonably large. Note that if we
only require the core to grow to $M_{\rm p}\approx 1.26M_{\oplus}$ (the
minimum core mass required if we assume $\kappa_{\rm d}=0.02$
cm$^{2}$/g), the density of solids should still be as large as
$\Sigma_{\rm d}\gtrsim 11 \mbox{g cm}^{-2}$, with little
improvement. In principle, another possible way to speed-up the core
accretion is to reduce the mass of the planetesimals. However, the
dependence on planetesimals mass is much shallower than that on surface
density [see Eq. (\ref{eq:coreacc})], so that even reducing $m$ by 10
orders of magnitude would only reduce the required surface density by
a factor of 10. 

We now consider the possible effect of planetary migration. When the
proto-planet has acquired enough mass to open up a gap in the disc, it
will undergo Type II migration, and its orbital evolution will be
therefore locked to the viscous evolution of the disc. The outer parts
of the disc spread out as a consequence of viscous forces, so, if the
protoplanet formed at a sufficiently large radius, it could in
principle migrate further out. We therefore proceed by estimating the
maximum radius at which the proto-planet could have reached
$4M_{\oplus}$ within 10 Myr, based on Eq. (\ref{eq:coreacc}), assuming
that $\Sigma_{\rm d}(R_{\rm s})=0.02$ g cm$^{-2}$, and considering the
radial dependence of $\Sigma\propto R^{-1}$. In this way, we find that
the maximum formation radius is $R_{\rm max}\approx 0.6$ AU. This
radius is well inside the typical radius separating the inward and
outward moving portions of the disc (which is typically 10 AU, see
\citealt{ida04}), so that even planetary migration would not reconcile
the core accretion model with the observed properties of the system.

Finally, note that the above estimates assume that the surface density
of solids and of gas in the disc is independent on time. In fact, both
solids and gas are depleted on a time-scale comparable to the life-time
of the disc. This would increase the core accretion time-scale and
would make it even harder [see Eq. (\ref{eq:coreacc})] for the
planetary core to grow to the required mass at late stages.

We can therefore conclude that the secondary in the system 2M1207
cannot have formed through core accretion, mainly because of its
distance from the primary, which leads to a very small surface density
of solids (even if the total disc mass is a significant fraction of the
primary mass) and to a very large core accretion time-scale. We note
that \citet{laughlin04} came to similar conclusions regarding the
possibility of forming giant planets through core accretion around M
dwarfs.

\subsection{Formation via gravitational instability}

The discussion of the previous section shows that there must be a
physical process {\it different from core accretion} able to form
planetary mass companions, at least around brown dwarfs. In the context
of planet formation, the natural alternative to the core accretion
model is the gravitational instability scenario, where planets form
from the fragmentation of a self-gravitating disc \citep{boss00}. Here
we examine the plausibility and the uncertainties of this model.

As we have already shown in the previous section, the total mass of the
disc $M_{\rm disc}$ must have been at least of the order of
$qM_{\star}=0.2M_{\star}$. Such a massive disc must have been
self-gravitating. In particular, the condition for being subject to
gravitational instability at radius $R$ (usually described in terms of
Toomre's stability criterion $Q=c_{\rm s}\kappa/\pi G\Sigma_{\rm
g}\approx 1$, where $c_{\rm s}$ is the sound speed and
$\kappa\approx\Omega$ is the epicyclic frequency) is that the disc
mass contained within $R$ satisfies the following relation:
\begin{equation}
\frac{M_{\rm disc}(R)}{M_{\star}}\gtrsim \frac{H}{R},
\label{eq:toomre}
\end{equation}
where $H$ is the disc thickness. In the following we will assume that
$H/R\approx 0.1$, a typical requirement for protostellar discs. What
would be the mass of a fragment produced in such an unstable disc? It
can be easily shown that the most gravitationally unstable wavelength
is $\lambda\approx 2\pi H$ and therefore the typical mass of a fragment
would be:
\begin{eqnarray}
\nonumber M_{\rm frag}&\approx& \Sigma_{\rm g}\lambda^2
=(2\pi)^2\Sigma_{\rm g}H^2\\
&=&2\pi\left(\frac{H}{R_{\rm s}}\right)^2M_{\rm disc}(R_{\rm s})\gtrsim
2\pi\left(\frac{H}{R_{\rm s}}\right)^3M_{\star},
\end{eqnarray}
where we have used the constraint that the disc mass at $R_{\rm s}$
should be high enough to be gravitationally unstable
[Eq. (\ref{eq:toomre})]. Putting in the relevant estimates for $H/R$
and $M_{\star}$, we get $M_{\rm frag}\approx 50M_{\oplus}$. Note that
this mass is large enough [cf. Eq. (1)] for the corresponding
Kelvin-Helmholtz time-scale to be shorter than the life-time of the
system. Therefore, if sufficiently resupplied, the fragment is able to
adjust its internal structure and grow to planetary mass. The mass
accretion rate needed to grow to $5M_{\rm Jup}$ in 10 Myrs is
$\dot{M}\approx 5\times 10^{-10}M_{\odot}/\mbox{yr}$. Accretion rates
of the order of $10^{-10} M_{\odot}/\mbox{yr}$ are usually observed in
the disc of brown dwarfs \citep{mohanty05}, and, since most of the
planet mass is accreted before the planet is able to open up a
substantial gap in the disc, it is able to accrete a sizable fraction
of the mass accretion rate through the disc.

The major obstacle in the gravitational instability scenario is that
the condition $Q\approx 1$ [or, equivalently, our Eq. (5)] is only a
necessary but not sufficient condition for fragmentation. The
development of a gravitational instability heats up the disc
\citep{LR04,LR05}, making it more stable. Fragmentation occurs only if
the cooling time-scale in the disc is sufficiently fast (of the order
of $3\Omega^{-1}$; \citealt{gammie01,rice03}). In fact, it has been
shown (Rice, Lodato \& Armitage 2005) that, depending on the assumed
adiabatic index in the disc, fragmentation can occur relatively more
easily, i.e. for cooling times of the order of $10\Omega^{-1}$. The
results of \citet{RLA05} suggest that gravitational instabilities
cannot provide an indefinitely large dissipation in the disc, larger
than $\alpha\approx 0.06$ (using the standard $\alpha$ description of
dissipative processes in discs). If this dissipation term is not
sufficient to balance the cooling of the disc, then fragmentation
occurs.

\citet{rafikov05} has computed the requirement on the disc structure in
order for the cooling to be fast enough to allow fragmentation, under
fairly general conditions. He found that fragmentation occurs if the
gas disc surface density is larger than a given threshold $\Sigma_{\rm
inf}$. If we compute $\Sigma_{\rm inf}$ from eq. (7) in
\citet{rafikov05}, and scale it down to our system, we find that
$\Sigma_{\rm inf}\approx 10 \mbox{g cm}^{-2}$ only slightly larger than
our simple estimate for the disc density $\Sigma_{\rm g}\approx
2.3~\mbox{g cm}^{-2}$. However, in his estimate of $\Sigma_{\rm inf}$,
\citet{rafikov05} assumed that the cooling time threshold for
fragmentation was $3\Omega^{-1}$, while, as discussed above,
\citet{RLA05} have shown that, depending on the adiabatic index, it can
be larger and up to $10\Omega^{-1}$. In this case, Eq. (7) of
\citet{rafikov05} would result in the smaller value of $\Sigma_{\rm
inf}\approx 6 \mbox{g cm}^{-2}$. We therefore conclude that disc
fragmentation is not unlikely for this particular system.

\subsection{Other binary formation mechanisms}

Binary stars have traditionally been thought to form via any of three
different mechanisms: capture, fission and fragmentation. Capture
supposes that the two components form independently and that, following
a close dissipative encounter, become bound. The probability for this
is remarkably low and thus this formation mechanism is highly unlikely
\citep{tohline02}. Fission assumes that a fast rotating contracting
protostar can reach the break up limit and split into two close
components \citep{durisen85}. There is no evidence that this can
actually happen in reality, although a final proof has not been given
yet \citep{tohline01}. However, we can discard this possibility, as
2M1207 is a relatively wide system, while this mechanism would
naturally produce close systems. Finally, binaries can form through
disc \citep{bonnell94} or core fragmentation
\citep{boss79,bonnell91}. Disc fragmentation has already been addressed
in the previous section. Core fragmentation involves some kind of $m=2$
perturbation in a collapsing cloud, which is later amplified by gravity
until the onset of Jeans instability which results in two individual,
albeit bound, pressure-supported objects. Gravitational fragmentation
in {\it isolated cores} can produce a large variety of binary systems,
depending on the size and mass of the core, and the amount of angular
momentum it contains \citep{bate00,fisher04}. The problem with this
scenario, from a theoretical point of view, is that simulations of star
formation with realistic initial conditions (e.g. random initial
`turbulent' velocities; \citealt{bate03,maclow04}) do not produce cores
that are as well defined as in models of isolated star formation. Cores
are seen as active entities, that grow in mass, that are typically
elongated because they are part of larger filamentary structures, and
that move in converging trajectories, driven by the underlying velocity
field and the gravitational attraction of high density regions.

Typically, star formation calculations
\citep{bate03,delgado04a,delgado04b,goodwin04b}
under-produce binary stars with low component masses. Extremely low
masses, such as those of the 2M1207 components, are only found among
objects that are promptly ejected from their parent cloud, and thus
cannot accrete much beyond the opacity limit for fragmentation
mass. The objects that are not ejected invariably grow to stellar
masses. However, ejection implies a bias against binarity. The binding
energy of 2M1207 is so low that it could have hardly survived as a
bound system after a close dynamical interaction with, by necessity, a
more massive binary. In this unlikely event, we would expect the system
to be eccentric. On the other hand, the primary could have been ejected
and the secondary formed in the circumprimary disc, which survived the
ejection process. This would bring us back to the gravitational
instability in a disc scenario (Section~3.2).

Current star formation calculations also under-produce binary stars
with mass ratios as low as that of 2M1207. As \citet{clarke05} have
shown, all processes that occur in a turbulent star forming cloud --
efficient fragmentation, intersecting flows, accretion of high angular
momentum material from circumbinary disc -- favour the formation of
bound pairs with similar masses from the start and, even for a binary
with initial low $q$, favour the evolution of the mass ratio towards
unity. However, these models cannot follow yet the long term evolution
of discs. Thus, in the context of dynamical star formation models
\citep{bate03,delgado04a,delgado04b,goodwin04b}, the
formation of systems such as 2M1207 is challenging, but not impossible,
provided that such systems turn out to be rare. 

\section{Conclusions}

In this paper we have discussed the nature and possible formation
mechanisms of 2M1207B, the $5M_{\rm Jup}$ companion of 2M1207A, a
$25M_{\rm Jup}$ brown dwarf. Even if the absolute mass of 2M1207B
places it clearly in the planetary range, we have discussed how, in
terms of separation and mass ratio, it is perfectly consistent as being
the secondary of a very low mass binary system. 

We have considered the formation mechanism of such system, and found
that the standard planet formation scenario of core accretion can be
ruled out, since the time-scale to form the planetary core is many
orders of magnitude larger than the observed age of the system. On the
other hand, we have shown that the alternative planet/binary formation
mechanism via gravitational instability leading to disc fragmentation
is a viable possibility in this particular case. In order for this
mechanism to work, however, the proto-brown dwarf must have been
surrounded by a massive disc in its early days, a rare (but possible)
event in the context of dynamical theories of star formation.

The very existence of 2M1207B poses therefore interesting constraints
on star formation theories. In fact, unless 2M1207 is a rare system,
that is, the result of an ejection of a brown dwarf with a large,
massive disc, its existence poses a challenge to the most dynamical
view of star formation. This system, if found not to be an exception,
seems to be saying that star formation can proceed sometimes more
``quietly'' (i.e. with fewer dynamical interactions) than seen in
current numerical simulations. In this sense, it is not surprising that
2M1207 has been found in the TW Hydrae association, a moving group
which was never massive enough to form a cluster, and hence is not
expected to have ever possessed a high enough stellar density for
widespread strong dynamical interactions to be commonplace. Thus, the
challenge from an observational point of view is not only to try to
constrain the occurrence of systems like 2M1207 but also their relative
frequency among different star forming environments. These observations
would provide interesting constraints on theories of star formation.

More generally, we have shown how the gravitational fragmentation of a
protostellar disc might be possible, at least under peculiar
circumstances (such as those that might have led to the formation of
2M1207). The main obstacles that work against this model for giant
planet formation around solar mass stars (long cooling times, high
fragment mass) are not a concern in this case. In fact,
\citet{rafikov05} has already shown that the cooling time problem is
not a serious concern at large distances ($\sim 100$ AU) from the
central star, such as the observed distance of 2M1207B from the
primary. Concerning the fragment mass, we have shown [Eq. (6)] that
$M_{\rm frag}\approx 2\pi (H/R)^3 M_{\star}$. Whereas for a solar mass
star the fragment mass is well in excess of Jupiter, for a brown dwarf
primary the fragment mass is much lower ($\sim 50 M_{\oplus}$) and
could lead to the formation of an object of planetary mass.

Finally, we note that the mechanism that we favour for the formation of
2M1207B (i.e. gravitational instability of a disc) is one that has been
invoked in the context of both binary formation and planet
formation. Whether 2M1207B should be classified as a ``binary
companion'' or a ``planet'' is therefore a semantic issue of secondary
interest.

\section*{Acknowledgement}

We thank Estelle Moraux for inspiring conversations and Simon Hodgkin
for a careful reading of the manuscript. CJC gratefully acknowledges
support from the Leverhulme Trust in the form of a Philip Leverhulme
Prize.

\bibliographystyle{mn2e} 
\bibliography{2mass}

\begin{thebibliography}{}

\bibitem[\protect\citeauthoryear{{Bate}}{{Bate}}{2000}]{bate00}
{Bate} M.~R.,  2000, MNRAS, 314, 33

\bibitem[\protect\citeauthoryear{{Bate}, {Bonnell} \& {Bromm}}{{Bate}
  et~al.}{2003}]{bate03}
{Bate} M.~R.,  {Bonnell} I.~A.,    {Bromm} V.,  2003, MNRAS, 339, 577

\bibitem[\protect\citeauthoryear{{Bonnell}, {Martel}, {Bastien}, {Arcoragi} \&
  {Benz}}{{Bonnell} et~al.}{1991}]{bonnell91}
{Bonnell} I.,  {Martel} H.,  {Bastien} P.,  {Arcoragi} J.-P.,    {Benz} W.,
  1991, ApJ, 377, 553

\bibitem[\protect\citeauthoryear{{Bonnell}}{{Bonnell}}{1994}]{bonnell94}
{Bonnell} I.~A.,  1994, MNRAS, 269, 837

\bibitem[\protect\citeauthoryear{Boss}{Boss}{2000}]{boss00}
Boss A.~P.,  2000, ApJ, 536, L101

\bibitem[\protect\citeauthoryear{{Boss} \& {Bodenheimer}}{{Boss} \&
  {Bodenheimer}}{1979}]{boss79}
{Boss} A.~P.,  {Bodenheimer} P.,  1979, ApJ, 234, 289

\bibitem[\protect\citeauthoryear{{Chauvin}, et~al.}{{Chauvin}
  et~al.}{2004}]{chauvin04}
{Chauvin} G., et al.,  2004, A\&A, 425, L29

\bibitem[\protect\citeauthoryear{{Chauvin}, et al.}{{Chauvin}
  et~al.}{2005}]{chauvin05}
{Chauvin} G.,  et al.,  2005, A\&A, 438, L25

\bibitem[\protect\citeauthoryear{{Clarke} \& {Delgado-Donate}}{{Clarke} \&
  {Delgado-Donate}}{2005}]{clarke05}
{Clarke} C.~J.,  {Delgado-Donate} E.~J.,  2005, in preparation

\bibitem[\protect\citeauthoryear{{Close}, {Siegler}, {Freed} \&
  {Biller}}{{Close} et~al.}{2003}]{close03}
{Close} L.~M.,  {Siegler} N.,  {Freed} M.,    {Biller} B.,  2003, ApJ, 587, 407

\bibitem[\protect\citeauthoryear{{Delgado-Donate}, {Clarke} \&
  {Bate}}{{Delgado-Donate} et~al.}{2004a}]{delgado04a}
{Delgado-Donate} E.~J.,  {Clarke} C.~J.,    {Bate} M.~R.,  2004a, MNRAS, 347,
  759

\bibitem[\protect\citeauthoryear{{Delgado-Donate}, {Clarke}, {Bate} \&
  {Hodgkin}}{{Delgado-Donate} et~al.}{2004b}]{delgado04b}
{Delgado-Donate} E.~J.,  {Clarke} C.~J.,  {Bate} M.~R.,    {Hodgkin} S.~T.,
  2004b, MNRAS, 351, 617

\bibitem[\protect\citeauthoryear{{Durisen} \& {Tohline}}{{Durisen} \&
  {Tohline}}{1985}]{durisen85}
{Durisen} R.~H.,  {Tohline} J.~E.,  1985, in Protostars and Planets II,
p.~534

\bibitem[\protect\citeauthoryear{{Fischer} \& {Marcy}}{{Fischer} \&
  {Marcy}}{1992}]{fischer92}
{Fischer} D.~A.,  {Marcy} G.~W.,  1992, ApJ, 396, 178

\bibitem[\protect\citeauthoryear{{Fisher}}{{Fisher}}{2004}]{fisher04}
{Fisher} R.~T.,  2004, ApJ, 600, 769

\bibitem[\protect\citeauthoryear{Gammie}{Gammie}{2001}]{gammie01}
Gammie C.~F.,  2001, ApJ, 553, 174

\bibitem[\protect\citeauthoryear{{Gizis}}{{Gizis}}{2002}]{gizis02}
{Gizis} J.~E.,  2002, ApJ, 575, 484

\bibitem[\protect\citeauthoryear{{Goodwin}, {Whitworth} \&
  {Ward-Thompson}}{{Goodwin} et~al.}{2004}]{goodwin04b}
{Goodwin} S.~P.,  {Whitworth} A.~P.,    {Ward-Thompson} D.,  2004, A\&A, 423,
  169

\bibitem[\protect\citeauthoryear{{Hubickyj}, {Bodenheimer} \&
  {Lissauer}}{{Hubickyj} et~al.}{2005}]{hubickyj05}
{Hubickyj} O.,  {Bodenheimer} P.,    {Lissauer} J.~J.,  2005, Icarus,
{\it in press}

\bibitem[\protect\citeauthoryear{{Ida} \& {Lin}}{{Ida} \& {Lin}}{2004}]{ida04}
{Ida} S.,  {Lin} D.~N.~C.,  2004, ApJ, 604, 388

\bibitem[\protect\citeauthoryear{{Klein}, {Apai}, {Pascucci}, {Henning} \&
  {Waters}}{{Klein} et~al.}{2003}]{klein03}
{Klein} R.,  {Apai} D.,  {Pascucci} I.,  {Henning} T.,    {Waters} L.~B.~F.~M.,
   2003, ApJ, 593, L57

\bibitem[\protect\citeauthoryear{{Laughlin}, {Bodenheimer} \&
  {Adams}}{{Laughlin} et~al.}{2004}]{laughlin04}
{Laughlin} G.,  {Bodenheimer} P.,    {Adams} F.~C.,  2004, ApJ, 612, L73

\bibitem[\protect\citeauthoryear{Lodato \& Rice}{Lodato \& Rice}{2004}]{LR04}
Lodato G.,  Rice W. K.~M.,  2004, MNRAS, 351, 630

\bibitem[\protect\citeauthoryear{Lodato \& Rice}{Lodato \& Rice}{2005}]{LR05}
Lodato G.,  Rice W. K.~M.,  2005, MNRAS, 358, 1489

\bibitem[\protect\citeauthoryear{{Luhman}}{{Luhman}}{2004}]{luhman04}
{Luhman} K.~L.,  2004, ApJ, 614, 398

\bibitem[\protect\citeauthoryear{{Mac Low} \& {Klessen}}{{Mac Low} \&
  {Klessen}}{2004}]{maclow04}
{Mac Low} M.-M.,  {Klessen} R.~S.,  2004, Reviews of Modern Physics, 76, 125

\bibitem[\protect\citeauthoryear{{Mazeh} \& {Goldberg}}{{Mazeh} \&
  {Goldberg}}{1992}]{mazeh92}
{Mazeh} T.,  {Goldberg} D.,  1992, ApJ, 394, 592

\bibitem[\protect\citeauthoryear{{Mohanty}, {Jayawardhana} \&
  {Basri}}{{Mohanty} et~al.}{2005}]{mohanty05}
{Mohanty} S.,  {Jayawardhana} R.,    {Basri} G.,  2005, ApJ, 626, 498

\bibitem[\protect\citeauthoryear{{Phan-Bao}, {Mart{\'{\i}}n}, {Reyl{\' e}},
  {Forveille} \& {Lim}}{{Phan-Bao} et~al.}{2005}]{phanbao05}
{Phan-Bao} N.,  {Mart{\'{\i}}n} E.~L.,  {Reyl{\' e}} C.,  {Forveille} T.,
  {Lim} J.,  2005, A\&A, 439, L19

\bibitem[\protect\citeauthoryear{Pollack et~al.,}{Pollack
  et~al.}{1996}]{pollack96}
Pollack J.~B.,  et~al., 1996, Icarus, 124, 62

\bibitem[\protect\citeauthoryear{{Rafikov}}{{Rafikov}}{2005}]{rafikov05}
{Rafikov} R.~R.,  2005, ApJ, 621, L69

\bibitem[\protect\citeauthoryear{Rice, Armitage, Bate \& Bonnell}{Rice
  et~al.}{2003}]{rice03}
Rice W. K.~M.,  Armitage P.~J.,  Bate M.~R.,    Bonnell I.~A.,  2003, MNRAS,
  338, 227

\bibitem[\protect\citeauthoryear{Rice, Lodato \& Armitage}{Rice
  et~al.}{2005}]{RLA05}
Rice W. K.~M.,  Lodato G.,    Armitage P.~J.,  2005, MNRAS, {\it in press}

\bibitem[\protect\citeauthoryear{{Sterzik}, {Pascucci}, {Apai}, {van der Bliek}
  \& {Dullemond}}{{Sterzik} et~al.}{2004}]{sterzik04}
{Sterzik} M.~F.,  {Pascucci} I.,  {Apai} D.,  {van der Bliek} N.,
  {Dullemond} C.~P.,  2004, A\&A, 427, 245

\bibitem[\protect\citeauthoryear{{Tohline}}{{Tohline}}{2002}]{tohline02}
{Tohline} J.~E.,  2002, ARA\&A, 40, 349

\bibitem[\protect\citeauthoryear{{Tohline} \& {Durisen}}{{Tohline} \&
  {Durisen}}{2001}]{tohline01}
{Tohline} J.~E.,  {Durisen} R.~H.,  2001, in IAU Symposium {An Update on Binary
  Formation by Rotational Fission}.
p.~40

\end{thebibliography}

\end{document}